	\documentclass[journal abbreviation]{copernicus}

\usepackage{graphicx}

\newcommand{\etal}{et al.}
\newcommand{\ie}{i.e.}
\newcommand{\be}{\begin{equation}}
\newcommand{\ee}{\end{equation}}
\newcommand{\beq}{\begin{eqnarray}}
\newcommand{\eeq}{\end{eqnarray}}

\newcommand{\aap}{    {\it Astron. Astrophys.}}

\newcommand{\apj}{    {\it Astrophys. J.}}
\newcommand{\apjl}{   {\it Astrophys. J. Lett.}}

\newcommand{\mnras}{  {\it Mon. Not. Roy. Astron. Soc.}}

\newcommand{\pasj}{   {\it Pub. Astron. Soc. Japan}}

\newcommand{\solphys}{{\it Solar Phys.}}
 
\newcommand{\ssr}{    {\it Space Sci. Rev.}}

\begin{document}
\title{Photospheric high-frequency acoustic power excess in sunspot umbra: signature of magneto-acoustic modes}
\author[1]{S.Zharkov}
\author[2]{S.Shelyag}
\author[3,4]{V.Fedun}
\author[4]{R.Erd{\'e}lyi}
\author[4,5]{M.J.Thompson}

\affil[1]{Department of Physics and Mathematics, University of Hull, Cottingham Road, Kingston-upon-Hull, HU6 7RX, UK }
\affil[2]{Monash Centre for Astrophysics,
School of Mathematical Sciences,
Monash University,
Clayton,
Victoria, AUSTRALIA 3800}
\affil[3]{Department of Automatic Control and Systems Engineering,
The University of Sheffield,
Mappin Street,
Sheffield, S1 3JD, 
UK}
\affil[4]{SP$^2$RC, School of Mathematics and Statistics,
          University of Sheffield, 
          Hounsfield Road, Hicks  
          Building, Sheffield, S3 7RH, UK}
\affil[5]{High Altitude Observatory, P.O. Box 3000,
Boulder, Colorado 80307-3000}

\runningtitle{Photospheric magneto-acoustic waves in sunspots}

\runningauthor{S.Zharkov \etal}

\correspondence{S. Zharkov\\ (s.zharov [at] hull.ac.uk)}

\received{}
\pubdiscuss{} 
\revised{}
\accepted{}
\published{}


\firstpage{1}

\maketitle  
\abstract{
We present  observational evidence for the presence of MHD waves in the solar 
photosphere deduced from SOHO/MDI Dopplergram velocity observations.  
The magneto-acoustic perturbations are observed as acoustic power 
enhancement in the sunspot umbra at high-frequency bands in the velocity component 
perpendicular to the magnetic field. 
We use numerical modelling of wave propagation through localised non-uniform
magnetic field concentration along with the same filtering procedure as applied to the observations 
to identify the observed waves. 
Guided by the results of the numerical simulations we
classify the observed oscillations as magneto-acoustic waves excited by the 
trapped sub-photospheric acoustic waves.
We consider the potential application of the presented method
as a diagnostic tool for magnetohelioseismology.
 \keywords{Solar: photosphere, sunspots --- helioseismology: acoustic oscillations, 
MHD, simulations}}

\introduction
Helioseismology, the study of acoustic oscillations excited by turbulence in the convection zone of the Sun, has been hugely successful in developing and testing our theories and models of solar interior \citep{duvall}.  Investigations 
of sunspots using methods of local helioseismology \citep{Kosovichev, zhao, zharkov07, thompsonzharkov, G2009}, 
which analyse the properties of waves passing through these magnetic features, 
have provided us with a wealth of insight into their subphotospheric nature and, in many cases, have posed new questions concerning the understanding of sunspot structure and its interaction with waves. 

So far such studies have mostly concentrated on the effect of sunspot magnetic structure on acoustic oscillations present in the quiet Sun photosphere. At the same time, it is known 
from MHD theory that a number of various oscillatory modes are present in 
magnetised atmospheres \citep{PE2011}. In fact, it is argued \citep{cally2, moradi1} that, at least some 
of, the inconsistencies in the helioseismic analyses of sunspots \citep{G2009} are likely due to not taking these modes into account. 
Numerical MHD simulations are currently used to help us gain an insight into such 
{problems} \citep{cally1, shelyag2, Shelyagb, shelyag3, parchevskii1, cameron, khomenko, felipe2010}. A variety of propagating 
and standing MHD waves (e.g., slow mode, Alfv{\'e}n, and fast mode) have been observed 
higher in the outer atmosphere of the Sun, mainly in coronal loops, but also in other 
structures such as coronal plumes and prominences \citep{bogdan, nagashima, fedun, J2009, ZE2009, MEJM2011, MVJ2012}. In addition, numerous observations 
obtained in {various} spectral lines revealed presence of three minute umbral oscillations from the transition region into the corona {\citep[e.g.][]{centeno, ban2007, sych2012}}. However, limited observational 
evidence of magneto-acoustic waves has been found so far at the level of the solar photosphere \citep{zirin, dorot, MEJM2011,MVFSE2013}. 

Based on preliminary analysis of symbiosis of observational and simulated data, 
in this paper we present a first direct evidence of presence of magneto-acoustic 
waves in the sunspot umbra. 
 Comparing the results of MHD forward modelling with the 
observations we show that the umbral power increase at high frequencies seen at large angles between the normal to 
the solar surface and the line of sight is consistent with the slow high-$\beta$ and fast low-$\beta$ waves.

\begin{figure}
\includegraphics{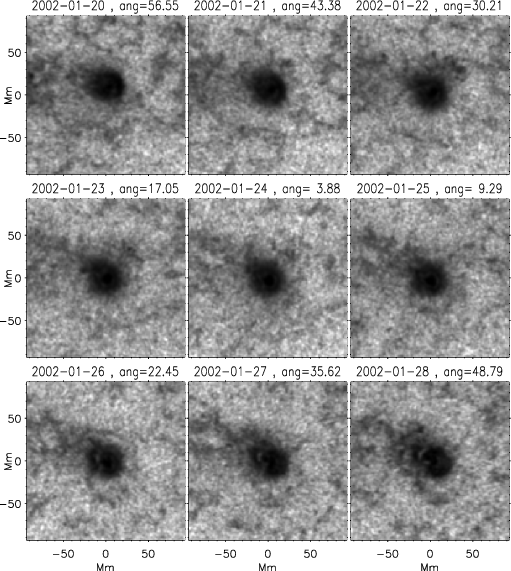}
\caption{
Acoustic power estimated from the line-of-sight velocity data for the NOAA 
AR9787 region. The data are taken at 9 subsequent snapshots as the sunspot travels from 
$56^\circ$ West to $61^\circ$ East on the solar surface. The data are frequency-filtered 
with the filter frequency band centered on 3 mHz.  
The images show little apparent variation in the centre of the sunspot.
\label{fig:obs_apower3mHz}}
\end{figure}
\begin{figure}
\includegraphics{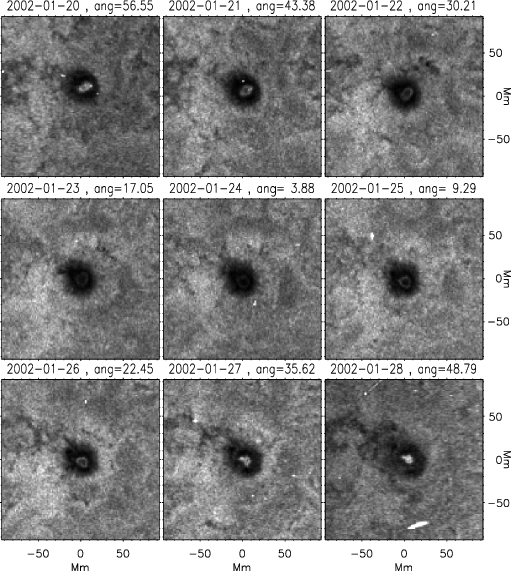}
\caption{
As Figure \ref{fig:obs_apower3mHz} but for the frequency band centered 6 mHz.
The images show a significant variation dependent on the heliographic 
angle and reveal easily noticeable power enhancement inside the sunspot at large angles 
from the solar disk centre (first and last frames of the panel). 
\label{fig:obs_apower6mHz}}
\end{figure}

\section{Data and Reduction}
\label{section1}
We investigate data from NOAA Active Region 9787, consisting of nearly single axisymmetric 
sunspot that showed little evolution during 20\,-\,28 January 2002, and observed 
continuously by the SOHO Michelson Doppler Imager (MDI) instrument. 
MDI uses the spectral line Ni-I 6776.772 \AA \ originating at approximately 300 km height 
above the solar surface \citep{scher_mdi}.
 The dataset, available on the European Helio- and 
Asteroseismology Network (HELAS) web site at \url{http://www.mps.mpg.de/projects/seismo/NA4/DATA/data_access.html}, 
was prepared by HELAS and has 
been thoroughly described and analysed by \cite{G2009}. The images 
were remapped using Postel projection with a map scale of $0.12^\circ$ to one 
$512\times512\times1440$ data cube of Doppler velocity data for each day. The centres 
of projection were chosen to track the motion of the sunspot (Carrington longitude of 
$\approx 133^\circ$ and latitude at $8.3^\circ$ South). 
Over the nine days of observations the region travelled from $56^\circ$ West to 
$61^\circ$ East. 

For each day of observation we compute the temporal Fourier transform of the 
Doppler images. We then divide this into 1-mHz bandwidth intervals and estimate 
the oscillatory power averaged over each of these frequency bandwidths. The power 
is then normalised by the dominating quiet Sun acoustic power value. Figures 
\ref{fig:obs_apower3mHz} and \ref{fig:obs_apower6mHz} show the results for the 3 mHz and 6 mHz centred frequency 
bands, respectively.

\section{Simulation}

\begin{figure*}
\includegraphics{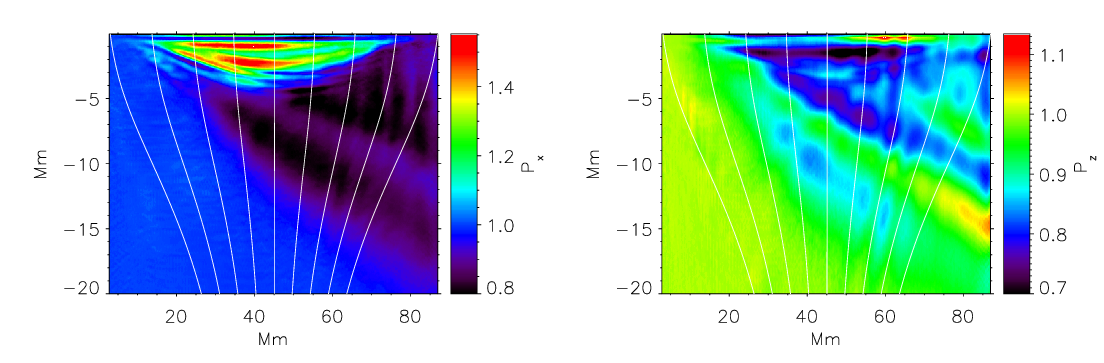}
\caption{Horizontal (left panel) and vertical (right panel) acoustic power ratios between 
the quiet and magnetic parts of the synthetic data obtained by means of forward modelling. The magnetic field lines are overplotted. Strong absorption is observed in the vertical acoustic power in the magnetised region, while
the horizontal power component shows a strong enhancement. Note that  {the scaling is different in vertical and horizontal directions is different}, 
thus the magnetic field lines in the plots do not show the actual inclination of the magnetic 
field in the model.  
\label{fig:sim_apower}}
\end{figure*}

We applied the code SAC (Sheffield Advanced Code) to carry out the simulations
of sound wave propagation through a localised strong non-uniform magnetic 
field concentration, representing a sunspot. A detailed description of the
code, numerical methods and the tests to show the robustness 
and applicability of SAC to a wide variety of magneto-hydrodynamic problems 
are presented {by} \citet{shelyag1}.

The code solves the full compressible system of MHD equations in  three, $(x, y, z)$ with $z$-axis in the vertical direction, or two, $(x, z)$ with all variables independent of $y$, 
dimensions on Cartesian grid. 
Hyperdiffusivity and hyperresistivity techniques
are used to ensure the numerical solution is stable {\citep{nord95}}. The code also uses variable 
separation to conserve the magneto-hydrostatic equilibrium of the background
unperturbed state. 

Standard Model S \citep{cdmodel}, slightly modified to achieve convective
stability, is implemented as the unperturbed ''quiet'', non-magnetic model of the solar interior.
The physical size of the computational domain is 180 Mm in the horizontal and 
50 Mm in the vertical directions, respectively. The domain is resolved by 960x1000 grid cells.
The numerical domain is set such that the upper boundary is located right
above the visible solar surface.

The boundaries of the domain are open, allowing the plasma to move into and out
of the numerical domain freely. However, some weak reflection {of the waves} from the boundaries
is observed due to {the not-ideal numerical representation} of the boundary conditions.

A non-uniform non-potential self-similar static magnetic field configuration 
\citep[see][]{ST1958, deinzer65, SchuRemp2005, shelyag3, Shelyagb, fse2011, fsvm2011, fvje2011} 
is implemented in half of the domain to mimic sunspot 
properties. The maximum vertical magnetic field strength is 3.5 kG at the level 
approximately corresponding to the visible solar surface. The magnetic field of 
this strength not only decreases the temperature in the sunspot at the solar
surface, but also creates a layer with the ratio of local Alfv{\'e}n speed to 
local sound speed greater than unity.  This layer may be responsible for magneto-acoustic wave mode conversion.

We use a single spatially and temporally localised acoustic source to excite
the oscillations in the domain. The source is located in the middle of the
horizontal layer 500 km beneath the solar surface. 
Due to symmetry of the source in the $x$-direction, this allows us to  directly compare the character of the wave propagation in the magnetic and non-magnetic halves of the domain.

\begin{figure*}
\includegraphics[width=8.5cm]{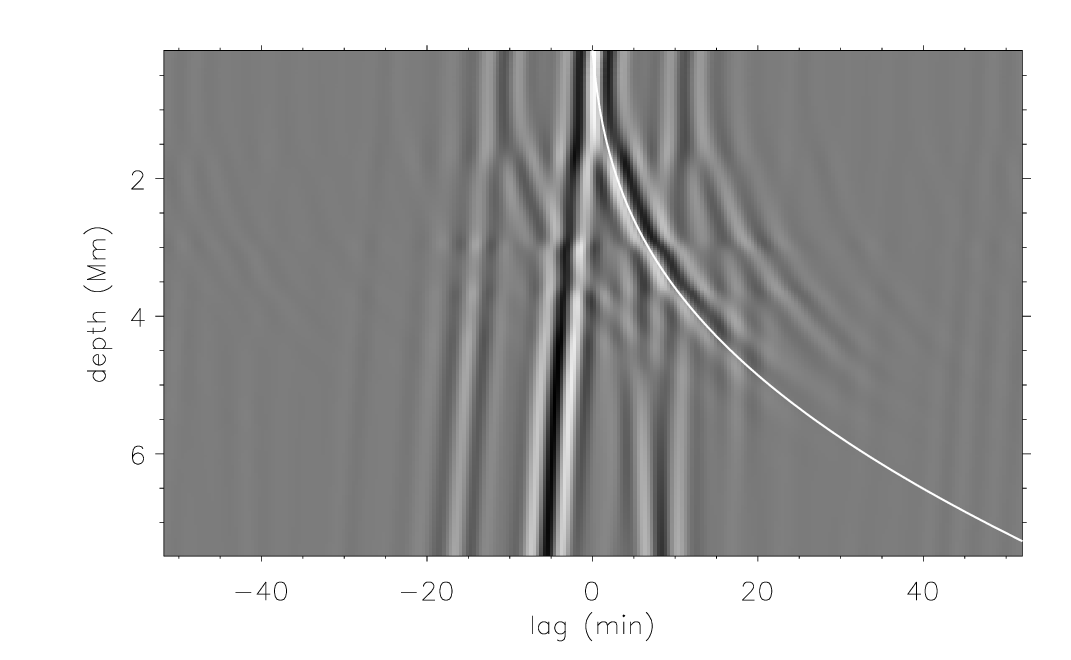}
\includegraphics[width=8.5cm]{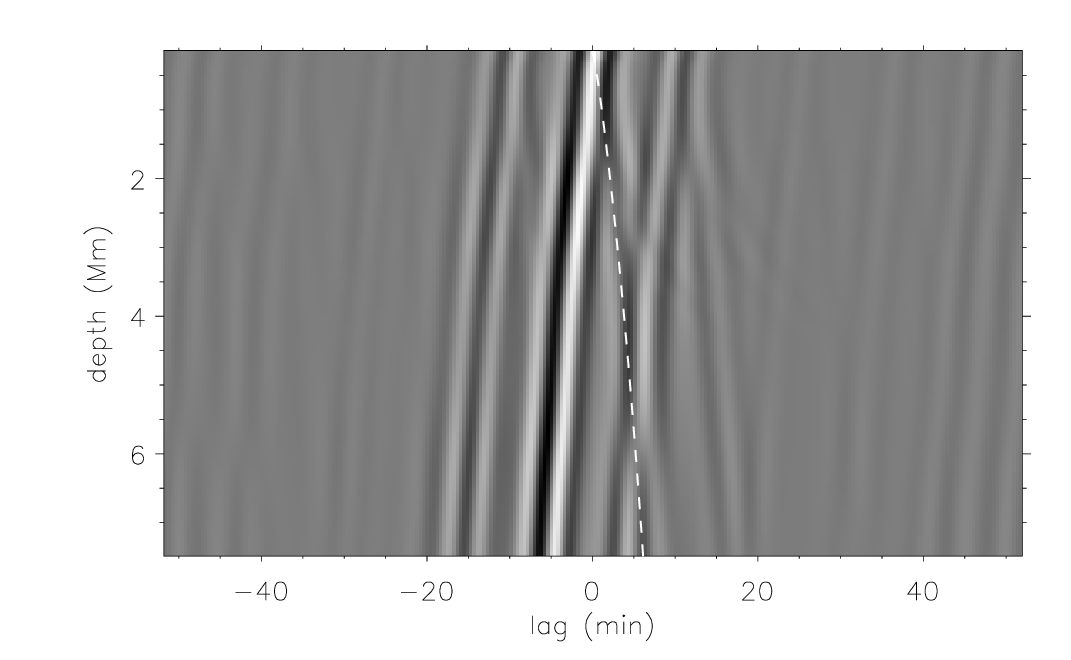}
\caption{Simulation cross-correlation in $v_x$ (left panel) and $v_z$ (right panel)
 at the center of the magnetic flux tube as a function of depth. 
The arriving fast magneto-acoustic wave and reflected slow wave can be seen in the $v_x$ cross-correlation, 
however, the fast wave dominates the $v_z$ cross-correlation image. The theoretical time-distance curve 
for the wave propagating with Alfv{\'e}n velocity is overplotted on the left panel (solid white line), and time-distance curve
 for the wave propagating with local sound speed is overplotted on the right one (dashed line). 
\label{fig:slow_mode_xc}}
\end{figure*}
\section{Results and Discussion}
The simulation used here was analysed using time-distance helioseismology, outlined in \citet{Shelyagb} as Case B where additional details about the magnetic configuration can be found.  The source generates acoustic wave-packet propagating through magnetic and non-magnetic halves of the simulation domain. When travelling through magentised plasma in 2D, the acoustic wave generally splits into slow and fast magneto-acoustic waves. 
However, in the high-$\beta$ region covering most of the computational domain except layers near the top, kinetic pressure dominates over magnetic. Thus, as shown in \citet{Shelyagb}, the fast high-$\beta$ wave generally dominates until it reaches the mode-conversion region, $v_A\approx v_s$ \ie \   $\beta \approx 1$, of the magnetic flux tube situated near the top of the domain. 
Then another mode becomes prominent in the magnetised domain, clearly seen in the computed magnetic field perturbation movies  which can be found at \url{http://robertus.staff.shef.ac.uk/publications/acoustic/}. Therefore the mode is magnetic in nature.

By considering the velocities in non-magnetic and magnetic parts of the same simulation we have 
observed that this mode is more pronounced in the horizontal rather than vertical 
component. This can be clearly seen in Figure \ref{fig:sim_apower}, where we have 
measured the acoustic power in the simulation box separately for horizontal and 
vertical velocity components,  $P_{x|z}(x, z)=\int v_{x|z}^2(x, z, t) dt$ and then constructed the acoustic power 
ratios between ''quiet'' and ''magnetic'' parts of the simulation as function of 
horizontal coordinate and depth. The acoustic power measured in the horizontal component of the velocity shows a pronounced enhancement in the near surface layers.

From theory, the wave group speed is related to the energy propagation. For the high-$\beta$ plasma slow mode wave it equals to the Alfv\'en speed and is almost parallel to magnetic field. For the high-$\beta$ fast the wave group speed is nearly independent of the magnetic field  \citep{Priest87,syr67}. 
To investigate these perturbations further we consider vertical propagation of the 
waves at the centre of the flux tube, $x_c,$ by cross-correlating the near surface velocity 
component measured at $z=z_{\rm surf}$ with deeper signal $C(z, t)=\int v_{x|z}(x_c, z_{\rm surf}, \tau) v_{x|z}(x_c, z, t+\tau) d\tau$. At the centre of the flux tube the magnetic field is exactly vertical by construction and the magnetic field line is parallel to the $z$-axis.
The results of cross-correlation are presented in Figure 
\ref{fig:slow_mode_xc} for the horizontal and vertical velocity components. For both components, 
we see the wave packet arriving to the surface from negative time-lag, being then reflected back to the interior for positive times. However, the correlation in the horizontal velocity component also shows the second, 
slower, mode propagating along the $z$-direction with the local Alfv{\'e}n speed. 
Thus, we conclude that, this is the slow-mode high-$\beta$ wave that propagates along the magnetic field lines generated by the mode-conversion of the original wave. 
Note that while this mode travels along the field line, directed in $z$-direction in this case, it is virtually absent in velocity $z$-component (right plot of Figure \ref{fig:slow_mode_xc}), which is 
in agreement with theory \cite{Priest87}. The counter-part of this mode above the mode-conversion region ($\beta<1$) is the fast low-$\beta$ mode, which also travels at Alfv{\'e}n speed and oscillates in the direction normal to it \citep{Priest87,KC2011,KC2012}. 





Thus we conclude that for almost vertical magnetic field, the slow high-$\beta$ mode will mainly contribute to the horizontal component of velocity. 
\begin{figure*}
\includegraphics[width=8.75cm]{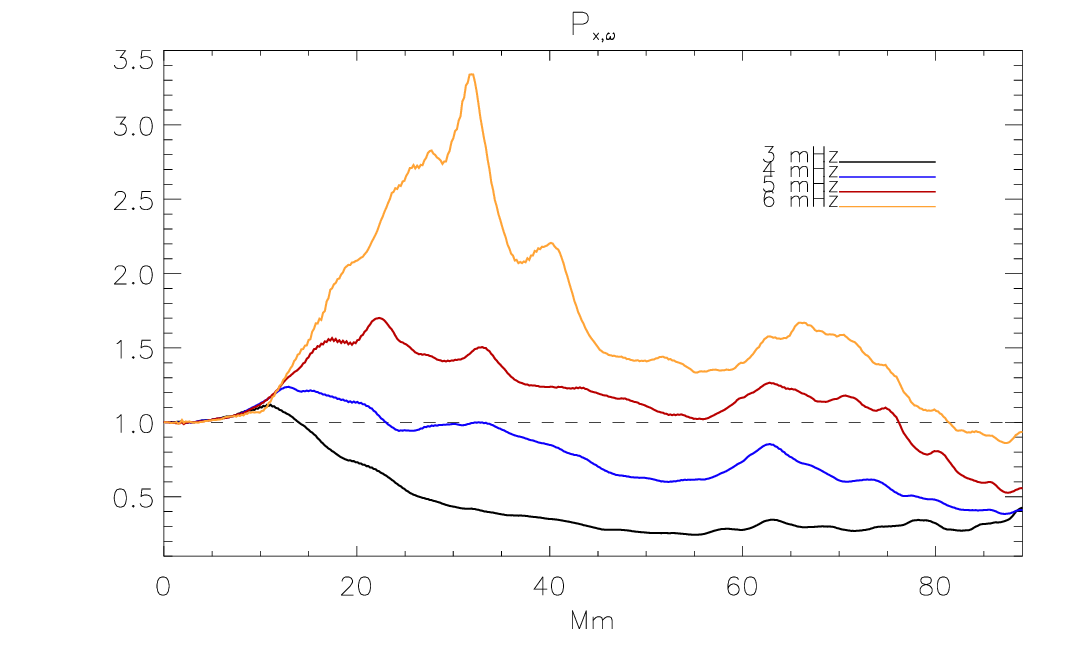}
\includegraphics[width=8.75cm]{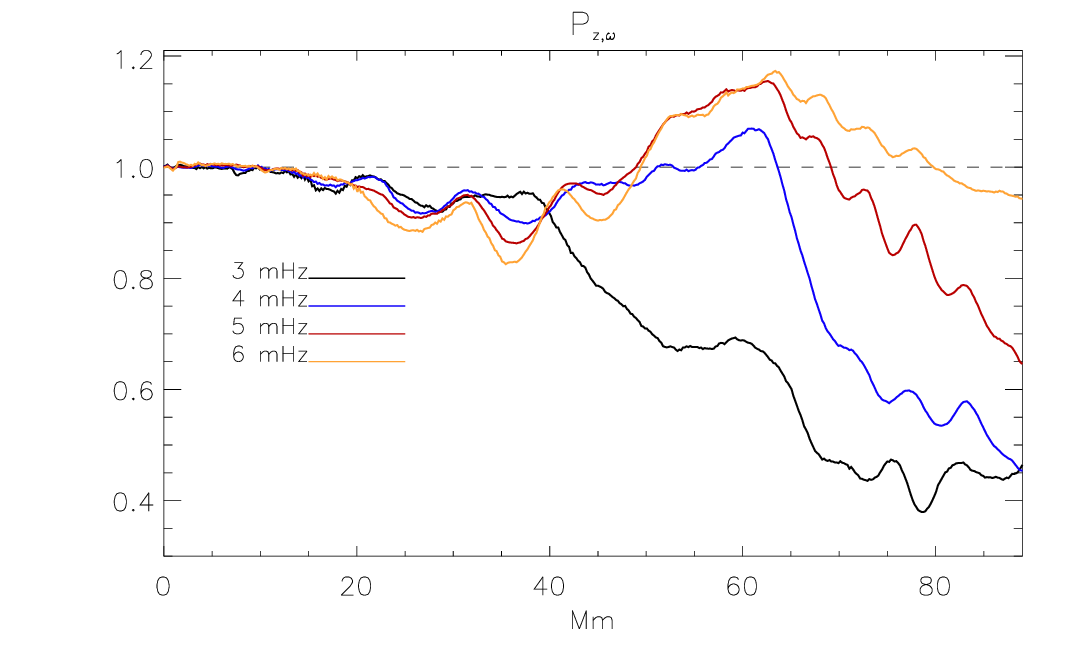}
\caption{The acoustic power ratio in horizontal (left plot) and vertical (right plot) velocity 
components in the simulation domain, filtered using the same filters as applied to observational data and described 
in Section \ref{section1}. A strong power enhancement at 6 mHz in horizontal velocity component
is observed, which corresponds to the power enhancement seen in the 6 mHz filtered observations 
(see Fig.~\ref{fig:obs_apower6mHz}). Again, no power enhancement is observed in the 3 mHz frequency band (cf. Fig.~\ref{fig:obs_apower3mHz}).
\label{fig:sim_freq_apower}}
\end{figure*}

In our 2D model we can describe the line-of-sight velocity $v_{\rm LOS}=v_z\cos \alpha +v_x\sin\alpha$, where $\alpha$ is the angle between the normal to the surface and line-of-sight direction. Clearly as $\alpha$ approaches $\pi/2$ the horizontal components dominates $v_{\rm LOS}$. In terms of solar observations, for a region located near the equator, $\alpha$ is approximately equal to the heliographic longitude of the observed region, therefore we can expect the stronger contribution of such magnetic modes at larger heliographic angles. 
The data in Figures  \ref{fig:obs_apower3mHz} and \ref{fig:obs_apower6mHz} show the acoustic power ratios 
in frequency bands centred at 3 and 6 mHz measured from daily NOAA 9787 
data from January 21 ($\alpha\approx 56^\circ$ West), when the region was 
located close to the west limb, to January 24 ($\alpha\approx 3^\circ$ West), 
when it was at the central meridian, to January 28 (close to east limb, 
$\alpha\approx 61^\circ$ East). At 6 mHz we can see the power enhancement (brightening)
at the central part of the sunspot, for the observations made close to the 
limb. No such effect is observed at 3 mHz. Also, ring-like structures of 
increased acoustic power are visible at the sunspot center at high 
frequencies at smaller heliographic angles, which suggests a possible 
signature of strong inclined fields. 

Such observational frequency dependence is also supported by our numerical simulation. Figure \ref{fig:sim_freq_apower} shows that the horizontal component oscillations inside the flux tube are strongly suppressed 
at 3 mHz, and enhanced at 6 mHz and above. Also in the vertical component the power increase at high frequencies is 
observed in the region where the magnetic field becomes slightly inclined. This could correspond to aforementioned ring like structures seen at high frequencies in observations close to the central meridian.

In order to verify our results, we consider the possible artifacts 
that might be present in the near limb data and can potentially affect 
the results of the analysis. SOHO/MDI are known to suffer from low light 
level ''saturation'' problem as well as from limitations of on-board 
processing algorithm. We have carefully considered the following 
potential artifacts: pixel rotation, low light levels and limitations 
of on-board processing algorithm \citep{technote}. Firstly, the pixel 
rotation does not cause a problem at the limb where the slow mode measurements 
are carried out. Also, following \cite{scher_ip} we have 
implemented the on-board algorithm, and using the intensity data from 
continuum observations, we have run several test simulations with varying 
noise levels. The results have shown little or no ill effect for acoustic power computation. 
Thus, we conclude that the acoustic power measurements derived from 
SOHO/MDI instrument are not affected by such issues.
\section{Conclusions}
We have found observational evidence for magneto-acoustic waves
deduced from high frequency acoustic power maps constructed from near-the-limb 
observations. Near-the-limb the observed Doppler shifted line-of-sight velocity signal has strong 
horizontal, \ie \  perpendicular to magnetic field, component. This behaviour 
agrees with theoretical description of the slow high-$\beta$ and fast low-$\beta$ modes. Although torsional Alfv\'en-type motions can be present in sunspots, here we were able to explain the excess power with two-dimensional numerical model where Alfv\'en waves are prohibited to propagate.

Supported by our simulation, we propose that in the high-$\beta$ interior most of the acoustic wave packet energy, when passing through the flux tube, goes into fast mode high-$\beta$ wave. This wave when travelling through the conversion region $v_A \approx v_s$, splits into fast and slow magneto-acoustic waves in our simulation. 
The fast magneto-acoustic high-$\beta$ wave passing through the tube interior transforms back to the acoustic mode, while the slow high-$\beta$ mode travels down along the field lines. 


 The fact that such slow oscillations are better pronounced in the direction 
normal to the magnetic field, opens a number of interesting questions and 
possibilities for helioseismic analysis of sunspot properties. This approach allows us to investigate the dependency between 
magnetic field angle at the surface, line of sight and oscillatory power, 
thus potentially providing information about magnetic field inclination. For example, our findings can be related to acoustic halo results reported in \citet{schunker2011}, where strong correlation of the enhanced high-frequency power with magnetic-field inclination was reported. 
It could also be of interest to model the effects of the acoustic sorce depth on geometry and interaction with magnetic field \cite{zharkov13, cally13}.

Finally, in this work we have used a HELAS dataset of over nine days of 
continuous observations of an isolated large and stable sunspot located at 
relatively low latitude. 
 It is the subject of future research 
to develop this analysis further and to extend it to other more extensive observations, for instance, using SDO data building on acoustic power analysis carried out in \cite{Rajaguru2012}. 

	\begin{acknowledgements}
	This work was initiated when all of the authors worked at School of Mathematics and Statistics, University of Sheffield.
	We would like to thank HELAS and for providing the data and support. 
	We also thank Tom Duvall, Phil Scherrer and Michael Ruderman
	for helpful discussions and guidance. RE acknowledges M. K\'eray for patient encouragement and is also grateful to NSF, Hungary (OTKA, Ref. Nos. K67746, K83133) for support received. SS research is supported by the Australian Research Council Future Fellowship.
	The authors would also like to thank the referees for their comments and input which have improved the paper.
	\end{acknowledgements}

\addtocounter{figure}{-1}\renewcommand{\thefigure}{\arabic{figure}a}


\begin{thebibliography}{}

\bibitem[Banerjee et al.(2007)]{ban2007} Banerjee, D., 
Erd{\'e}lyi, R., Oliver, R., \& O'Shea, E.\ 2007, \solphys, 246, 3 

\bibitem[Bogdan et al.(2003)]{bogdan} Bogdan, T.~J., et al.\ 
2003, \apj, 599, 626 

\bibitem[Cally(2013)]{cally13} Cally, P.~S.\ 2013, \apj, 768, 
35 

\bibitem[Cameron et al.(2008)]{cameron} Cameron, R., Gizon, L., 
\& Duvall, T.~L., Jr.\ 2008, \solphys, 51 

\bibitem[Centeno et al.(2006)]{centeno} Centeno, R., Collados, 
M., \& Trujillo Bueno, J.\ 2006, \apj, 640, 1153 

\bibitem[Crouch 
\& Cally(2003)]{cally1} Crouch, A.~D., \& Cally, P.~S.\ 2003, \solphys, 214, 201 

\bibitem[Christensen-Dalsgaard et al.(1996)]{cdmodel} 
Christensen-Dalsgaard, J., et al.\ 1996, Science, 272, 1286 

\bibitem[Deinzer(1965)]{deinzer65} Deinzer, W.\ 1965, \apj, 141, 
548 


\bibitem[Dorotovi{\v c} et al.(2008)]{dorot} Dorotovi{\v c}, 
I., Erd{\'e}lyi, R., \& Karlovsk{\'y}, V.\ 2008, IAU Symposium, 247, 351 

\bibitem[Duvall et al.(1997)]{duvall} Duvall, T.~L., Jr., et 
al.\ 1997, \solphys, 170, 63 


 \bibitem[Fedun et al.(2009)]{fedun} Fedun, V., Erd{\'e}lyi, 
R., \& Shelyag, S.\ 2009, \solphys, 258, 219

\bibitem[Fedun et al.(2011a)]{fse2011} Fedun, V., Shelyag, S., 
\& Erd{\'e}lyi, R.\ 2011, \apj, 727, 17

\bibitem[Fedun et al.(2011b)]{fsvm2011} Fedun, V., Shelyag, S., 
Verth, G., Mathioudakis, M., 
\& Erd{\'e}lyi, R.\ 2011, Annales Geophysicae, 29, 1029 

\bibitem[Fedun et al.(2011c)]{fvje2011} Fedun, V., Verth, G., 
Jess, D.~B., \& Erd{\'e}lyi, R.\ 2011, \apjl, 740, L46 

\bibitem[Felipe et al.(2010)]{felipe2010} Felipe, T., Khomenko, 
E., \& Collados, M.\ 2010, \apj, 719, 357

\bibitem[Gizon et al.(2009)]{G2009} Gizon, L., et al.\ 2009, 
Space Science Reviews, 144, 249 


\bibitem[Jess et al.(2009)]{J2009} Jess, D.~B., Mathioudakis, 
M., Erd{\'e}lyi, R., et al.\ 2009, Science, 323, 1582 


\bibitem[Khomenko 
\& Cally(2011)]{KC2011} Khomenko, E., \& Cally, P.~S.\ 2011, Journal of Physics Conference Series, 271, 012042 

\bibitem[Khomenko 
\& Cally(2012)]{KC2012} Khomenko, E., \& Cally, P.~S.\ 2012, \apj, 746, 68 


\bibitem[Khomenko et al.(2009)]{khomenko} Khomenko, E., 
Kosovichev, A., Collados, M., Parchevsky, K., 
\& Olshevsky, V.\ 2009, \apj, 694, 411 

\bibitem[Kosovichev 
\& Duvall(1997)]{Kosovichev} Kosovichev, A.~G., \& Duvall, T.~L., Jr.\ 1997, SCORe'96 : Solar Convection and Oscillations and their Relationship, 225, 241
	
\bibitem[Liu \& Norton(2001)]{technote} Liu, Y., \& Norton, A.~A.\ 2001, \url{http://soi.stanford.edu/technotes/01.144/TN01-144.pdf} 

\bibitem[Moradi 
\& Cally(2008)]{cally2} Moradi, H., \& Cally, P.~S.\ 2008, \solphys, 251, 309 

\bibitem[Moradi et al.(2009)]{moradi1} Moradi, H., Hanasoge, 
S.~M., \& Cally, P.~S.\ 2009, \apjl, 690, L72 


\bibitem[Morton et al.(2011)]{MEJM2011} Morton, R.~J., 
Erd{\'e}lyi, R., Jess, D.~B., \& Mathioudakis, M.\ 2011, \apjl, 729, L18 


\bibitem[Morton et al.(2013)]{MVFSE2013} Morton, R.~J., Verth, 
G., Fedun, V., Shelyag, S., \& Erd{\'e}lyi, R.\ 2013, \apj (accepted) arXiv:1303.2356 

\bibitem[Morton et al.(2012)]{MVJ2012} Morton, R.~J., Verth, 
G., Jess, D.~B., et al.\ 2012, Nature Communications, 3,  

\bibitem[Nagashima et al.(2007)]{nagashima} Nagashima, K., et 
al.\ 2007, \pasj, 59, 631 

\bibitem[Nordlund \& Galsgaard (1995)]{nord95} Nordlund, \r{A} and Galsgaard, K, 1995, Technical Report, Copenhagen: Astronomical Observatory

\bibitem[Parchevsky 
\& Kosovichev(2007)]{parchevskii1} Parchevsky, K.~V., \& Kosovichev, A.~G.\ 2007, \apj, 666, 547 

\bibitem[Pint{\'e}r 
\& Erd{\'e}lyi(2011)]{PE2011} Pint{\'e}r, B., \& Erd{\'e}lyi, R.\ 2011, \ssr, 158, 471 


\bibitem[Priest(1987)]{Priest87} Priest, E.~R.\ 1987, Solar 
magneto-hydrodynamics by E.R.~Priest.~Dordrecht: D.~Reidel, 1987.,  


\bibitem[Rajaguru et al.(2012)]{Rajaguru2012} Rajaguru, S.~P., 
Couvidat, S., Sun, X., Hayashi, K., \& Schunker, H.\ 2012, \solphys, 303 


\bibitem[Scherrer(1993)]{scher_ip} Scherrer, P.~H.\ 1993, \url{http://soi.stanford.edu/technotes/94.110.pdf} 

\bibitem[Scherrer et al.(1995)]{scher_mdi} Scherrer, P.~H., et al.\ 1995, \solphys, 162, 129 

\bibitem[Schunker 
\& Braun(2011)]{schunker2011} Schunker, H., \& Braun, D.~C.\ 2011, \solphys, 268, 349 

\bibitem[Sch{\"u}ssler 
\& Rempel(2005)]{SchuRemp2005} Sch{\"u}ssler, M., \& Rempel, M.\ 2005, \aap, 441, 337 

\bibitem[Schl{\"u}ter 
\& Temesv{\'a}ry(1958)]{ST1958}Schl{\"u}ter, A., \& Temesv{\'a}ry, S.\ 1958, Electromagnetic Phenomena in Cosmical Physics, 6, 263 

\bibitem[Shelyag et al.(2007)]{shelyag2} Shelyag, S., 
Erd{\'e}lyi, R., \& Thompson, M.~J.\ 2007, \aap, 469, 1101 

\bibitem[Shelyag et al.(2008)]{shelyag1} Shelyag, S., Fedun, V., 
\& Erd{\'e}lyi, R.\ 2008, \aap, 486, 655 
 
\bibitem[Shelyag et 
al.(2010)]{shelyag3} Shelyag, S., Mathioudakis, M., Keenan, F.~P., \& Jess, D.~B.\ 2010, \aap, 515, A107

\bibitem[Shelyag et al.(2009)]{Shelyagb} 
Shelyag, S., Zharkov, S., Fedun, V., Erd{\'e}lyi, R., \& Thompson, M.~J.\ 2009, \aap, 501, 735 

\bibitem[Sych et 
al.(2012)]{sych2012} Sych, R., Zaqarashvili, T.~V., Nakariakov, V.~M., et al.\ 2012, \aap, 539, A23 


\bibitem[Syrovatskii 
\& Zhugzhda(1967)]{syr67} Syrovatskii, S.~I., \& Zhugzhda, Y.~D.\ 1967, {\it Astrof. Zhunal}, 44, 1180 

\bibitem[Thompson 
\& Zharkov(2008)]{thompsonzharkov} Thompson, M.~J., \& Zharkov, S.\ 2008, \solphys, 251, 225 

\bibitem[Zhao 
\& Kosovichev(2006)]{zhao} Zhao, J., \& Kosovichev, A.~G.\ 2006, \apj, 643, 1317 

\bibitem[Zaqarashvili 
\& Erd{\'e}lyi(2009)]{ZE2009} Zaqarashvili, T.~V., \& Erd{\'e}lyi, R.\ 2009, \ssr, 149, 355 

\bibitem[Zharkov(2013)]{zharkov13} 
Zharkov, S., 2013, \mnras, 431, 3414

\bibitem[Zharkov et al.(2007)]{zharkov07} Zharkov, S., Nicholas, 
C.~J., \& Thompson, M.~J.\ 2007, Astronomische Nachrichten, 328, 240 

\bibitem[Zirin 
\& Stein(1972)]{zirin} Zirin, H., \& Stein, A.\ 1972, \apjl, 178, L85 

\end{thebibliography}
\end{document}